# Structural Phase Transition in the Superconducting Pyrochlore Oxide $Cd_2Re_2O_7$

M. Hanawa, J. Yamaura, Y. Muraoka, F. Sakai, Z. Hiroi

*Institute for Solid State Physics, University of Tokyo, Kashiwanoha, Kashiwa, Chiba, 277-8581, Japan*

## Abstract

We report a structural phase transition found at $T_s$ = 200 K in a pyrochlore oxide $Cd_2Re_2O_7$ which shows superconductivity at $T_c$ = 1.0 K. X-ray diffraction experiments indicate that the phase transition is of the second order, from a high-temperature phase with the ideal cubic pyrochlore structure (space group $Fd\bar{3}m$) to a low-temperature phase with another cubic structure (space group $F\bar{4}3m$). It is accompanied by a dramatic change in the resistivity and magnetic susceptibility and thus must induce a significant change in the electronic structure of $Cd_2Re_2O_7$.

## 1. Introduction

Pyrochlore oxides belong to one of the largest structural groups for transition metal (TM) oxides. They are constituted by the chemical formula $A_2B_2O_7$, where the A site is occupied by a rare-earth or a post-transitioin metal ion with eightfold oxygen coordination and the B site by a transition metal ion with sixfold oxygen coordination [1]. The A and B sites individually form corner-sharing tetrahedral sublattices which are inter-penetrating with each other. If either the A or B site is occupied by a magnetic atom with an antiferromagnetic nearest-neighbor interaction, a geometrical frustration should be present, which might give rise to an unusual ground state [2].

On the other hand, *d* electrons from TM ions often exhibit an itinerant character, depending on the combination of the two cations. Looking in the general trend of electronic properties for pyrochlore oxides, many of 3*d* and 4*d* pyrochlore oxides are insulators owing to large electron correlation as well as relatively small electron transfer along the bent B-O-B bonds. In contrast, 5*d* pyrochlore oxides often show a metallic conductivity due to direct overlap of 5*d* orbitals [1]. A rare exception reported previously is found in $Os^{5+}$ pyrochlores like $Cd_2Os_2O_7$ [3, 4], where a metal-insulator (MI) transition occurs at $T$ = 226 K in the absence of structural anomalies. An $Os^{5+}$ ion has a $5d^3$ electron configuration and thus the $t_{2g}$ orbital is half-filled, suggesting a possible Mott-Hubbard type MI transition. This example suggests that the effect of electron correlation is still important even for the 5*d* electron systems in the pyrochlore structure. It is intriguing to study how correlated electrons behave on the highly-frustrated pyrochlore lattice, but has not yet been explored intensively. It is to be noted that no superconductivity had been observed there so far in spite of many metallic compounds present in the family.

Very recently, we prepared single crystals of $Cd_2Re_2O_7$ containing a $Re^{5+}$ ion with a $5d^2$ electron configuration and found superconductivity at a critical temperature $T_c$ = 1.0 K [5]. Sakai *et al.* also reported superconductivity in the same compound independently [6], and Jin *et al.* reported similar results later [7]. The compound has been known to possess the ideal pyrochlore structure at room temperature, and show metallic conductivity down to $T$ = 4 K [8]. In addition to the observation of super-conductivity in $Cd_2Re_2O_7$ we found remarkable anomalies at 200 K both in the electrical resistivity and magnetic susceptibility measurements [5]. It was suggested that the anomalies indicated another phase transition, which induced a significant change in the electronic structure. Here we report a structural study by means of X-ray diffraction (XRD) on $Cd_2Re_2O_7$ which clearly reveals the presence of a second-order structural phase transition at $T_s$ = 200 K.

## 2. Experimental

Single crystals of $Cd_2Re_2O_7$ were prepared by solid state reaction, 2CdO + 5/3$ReO_3$ + 1/3Re → $Cd_2Re_2O_7$, in an evacuated silica tube at 900˚C for 72h. Purple octahedral crystals of a few mm in length were obtained. From the facts that both CdO (Cd) and $ReO_3$ ($Re_2O_7$) are volatile and that crystals adhered to the walls, it is considered that the crystal growth occurred by a vapor transport mechanism, as discussed in a previous study [8]. The Cd/Re molar ratio was 1.00 ± 0.02 as determined by electron-probe microanalyses (EPMA) and was 0.995 as determined by inductively coupled plasma (ICP) spectrometry. The oxidation state of Re was about 5.1, which was determined by oxidizing a sample with $Ce^{4+}$ and detecting the remnant oxidizer with iodometric titration.

A few single crystals were crushed into powder and examined by means of powder XRD. The pattern taken at room temperature was indexed on the basis of a face centered unit cell with $a$ = 1.0226(2) nm, which is slightly larger than the previously reported value (1.0219 nm) [8]. Low temperature experiments were carried out using a cryocooler down to $T$ = 10 K. The cell constant at each temperature was refined by the Rietveld method with a program RIETAN [9], using Si powder as an internal standard [10]. Single crystal XRD experiments were performed using the $2\theta$-$\omega$ scan mode in a four-circle diffractmeter with a gas flow cooler down to $T$ = 100 K. Specific heat and electrical resistivity were measured in a Quantum Design PPMS system equipped with a $^3$He refrigerator down to $T$ = 0.4 K. Magnetic susceptibility measurements were performed at temperatures between 1.7 and 700 K in a Quantum Design MPMS system.

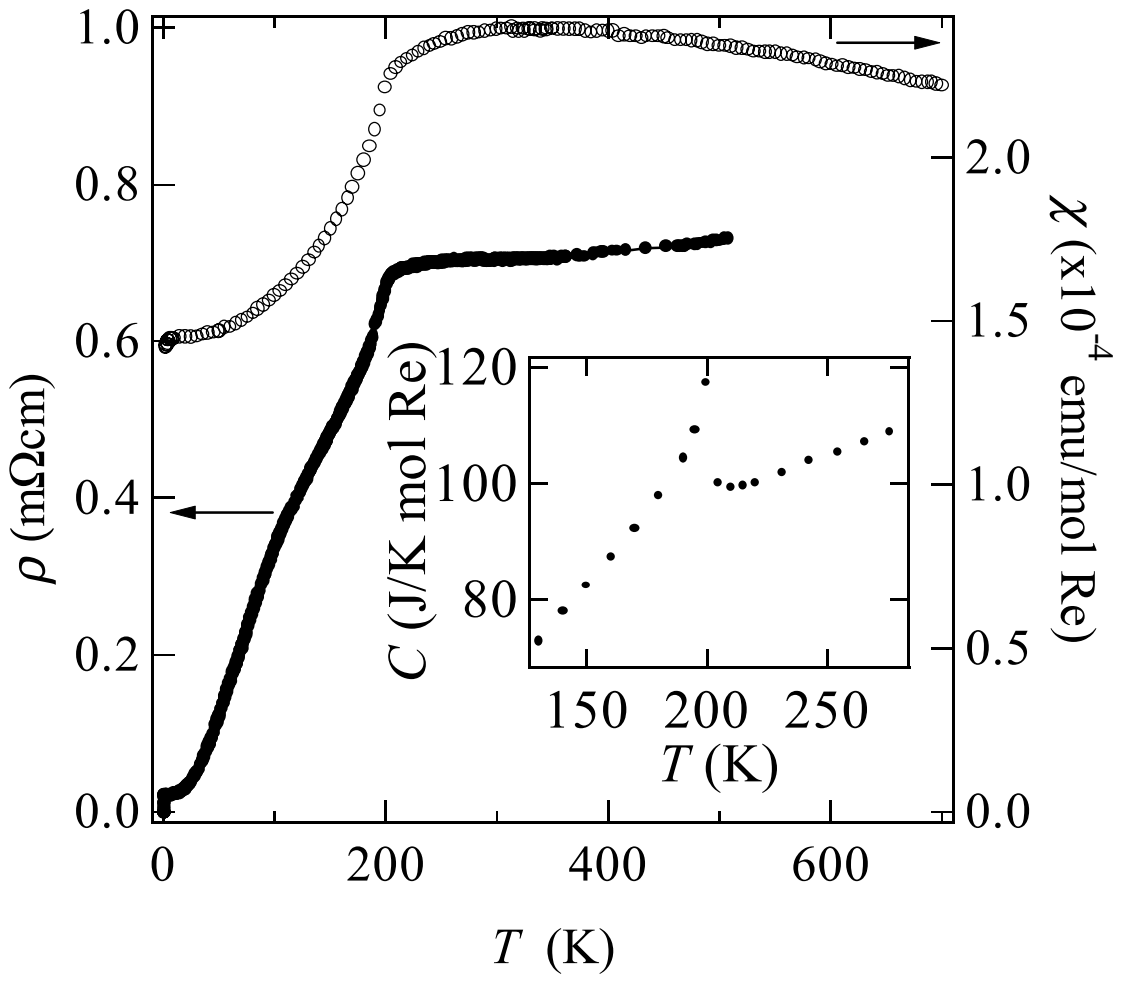

Fig. 1. Temperature dependence of electrical resistivity $\rho$ (left axis), magnetic susceptibility $\chi$ (right axis), and specific heat $C$ (inset) of a $Cd_2Re_2O_7$ single crystals, showing anomalies at $T_s$ = 200 K. The measurement of magnetic susceptibility was performed in a magnetic field of 1 T.

## 3. Results and Discussion

Anomalies observed at $T_s$ = 200 K in the resistivity $\rho$ and magnetic susceptibility $\chi$ are shown in Fig. 1, where the data above room temperature is added to the data previously reported [5]. The resistivity is almost temperature independent above $T_s$ up to 500 K, while steeply decreases below $T_s$. A superconducting transition was observed at $T_c$ ~ 1.0 K for this crystal. The magnetic susceptibility also decreases suddenly below $T_s$, reduced by about 30 % at $T$ = 5 K compared with the value at $T_s$. Preliminary NMR experiments by Takigawa *et al.* [11] indicated the absence of magnetic order. Above $T_s$, the $\chi$ curve exhibits a broad, rounded maximum around 320 K and decreases gradually on heating up to 700 K, which is far from the simple Pauli paramagnetic behavior. No anisotropy was detected in $\chi$ throughout measurements using polycrystalline samples and single crystals with magnetic fields applied along the <111> and <110> axes. Specific heat measurements gave a strong evidence for a thermodynamic phase transition at $T_s$, as shown in the inset to Fig. 1. The anomaly observed has a typical λ-shape, suggesting a second-order transition.

Next we have checked possibility for a structural phase transition. We did not detect splitting or broadening of certain peaks in the powder XRD pattern below $T_s$ = 200 K, implying that the crystal system of the low-temperature phase is still cubic. However, the temperature dependence of the cell constant $a$ exhibited a distinct knee at 200 K (Fig. 2): On cooling from room temperature, the cell constant decreases gradually, and starts to increase below $T_s$. For reference, we measured the temperature dependence of the cell constant for another pyrochlore oxide $Cd_2Os_2O_7$ in the same experimental way. As mentioned already, $Cd_2Os_2O_7$ exhibits a MI transition at $T$ = 226 K without any signs for structural changes [3, 4]. Its cell constant decreases monotonically on cooling in the whole temperature range, which may represent a substantial thermal expansion for an undistorted pyrochlore oxide. The linear coefficient of thermal expansion defined as $\alpha = \frac{1}{a}\left(\frac{\partial a}{\partial T}\right)$ is nearly the same at room temperature, $\alpha = 4.9\times10^{-6}$ $K^{-1}$ and $4.3\times10^{-6}$ $K^{-1}$ for the Re and Os compounds, respectively.

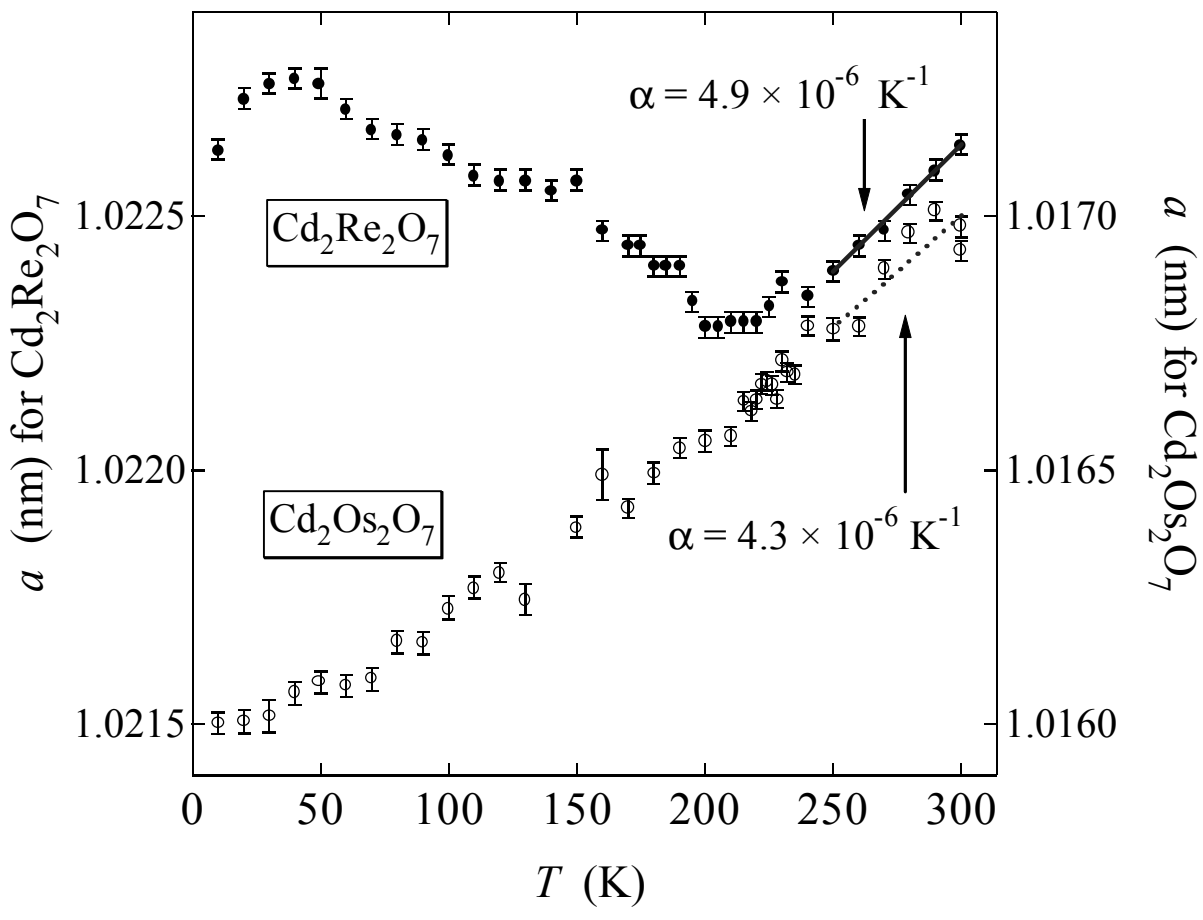

Fig. 2. Temperature dependence of the cell constant $a$ of $Cd_2Re_2O_7$ (left axis, filled circle) and $Cd_2Os_2O_7$ (right axis, open circle). The linear coefficient of thermal expansion $\alpha$ around room temperature is nearly same for the two compounds.

Since the low-temperature phase is still cubic, a change in the space group must occur at the transition. We carried out measurements using a single crystal on a four-circle diffractometer. Some extra reflections Appeared below $T_s$, which do not obey the extinction rule of space group $Fd\bar{3}m$. As displayed in Fig. 3, the intensity of the extra reflections grow gradually on cooling below $T_s$, suggesting a second-order transition.

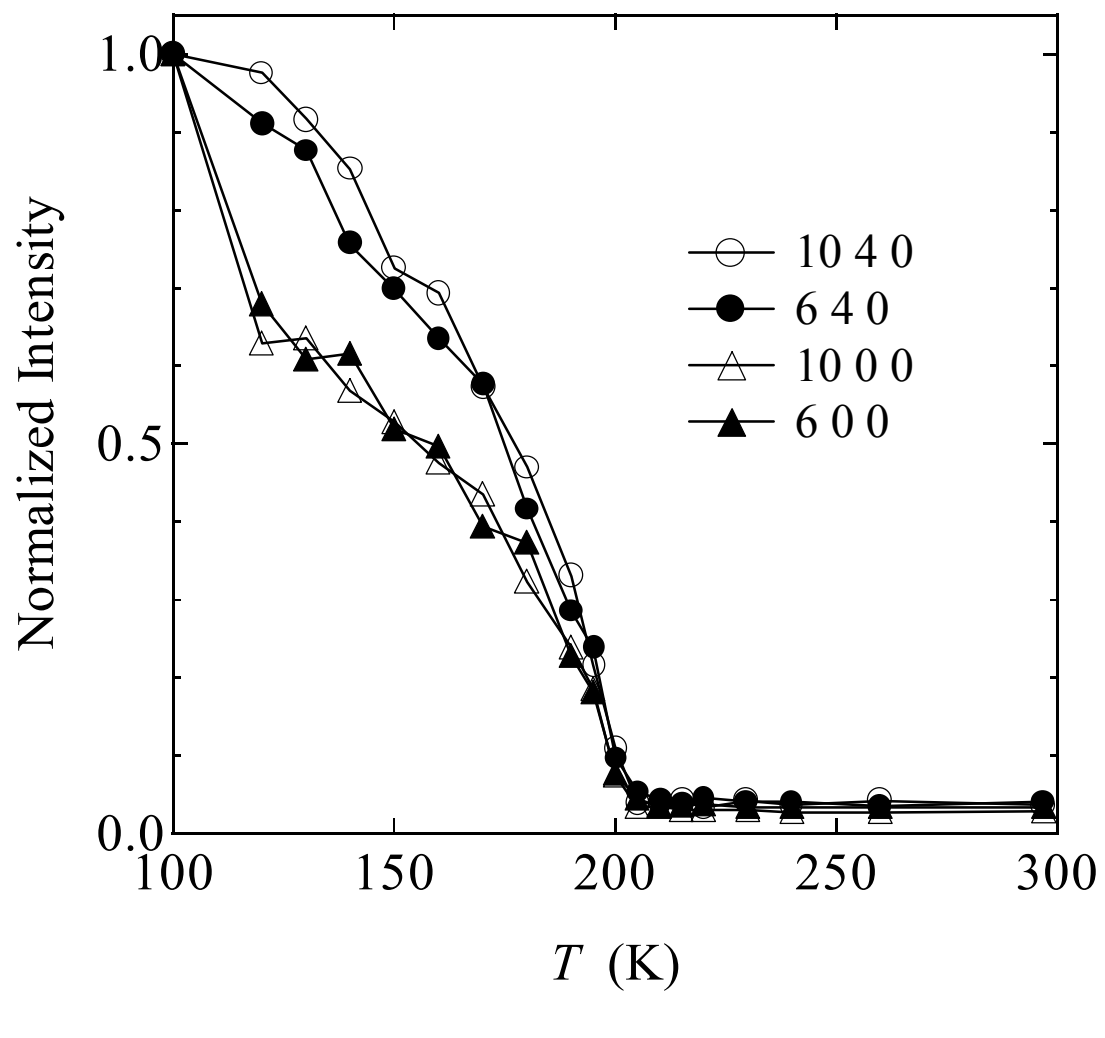

Fig. 3. Temperature dependence of the XRD intensity of extra reflections normalized at 100 K. All the reflections displayed do not obey the extinction rule of space group $Fd\bar{3}m$ for the high-temperature phase.

Examination of equivalent reflections also indicated that $Cd_2Re_2O_7$ retained the cubic symmetry ($m\bar{3}m$) throughout the transition. Assuming a second-order phase transition, possible cubic space groups for the low-temperature phase are $Fd\bar{3}$, $F4_132$ and $F\bar{4}3m$ which belong to the subgroups of $Fd\bar{3}m$. We have finally selected space group $F\bar{4}3m$ to explain the observed systematic absence of reflections at 100 K. For example, the low-temperature phase exhibited the (600) reflection, which did not agree with the reflection conditions of space group $Fd\bar{3}$ or $F4_132$ ($h$00: $h$ = 4$n$). Very recently we confirmed this structural change with preliminary neutron diffraction experiments using polycrystalline sample of $^{114}Cd_2Re_2O_7$, which will be reported elsewhere [12]. It is to be noted that an oxygen-deficient pyrochlore oxide $Pb_2Ru_2O_{6.5}$ possesses the same space group $F\bar{4}3m$ at room temperature which is believe to be induced by the ordering of oxygen atoms and vacancies [13].

It is interesting to notice that the symmetric position (notation with multiplicity and Wyckoff letter) of Re atoms changes from 16$c$ (1/8, 1/8, 1/8) in $Fd\bar{3}m$ (origin choice 1) to 16$e$ ($x$, $x$, $x$) in $F\bar{4}3m$. This means that Re atoms can displace along the <111> direction in the low-temperature phase. Consequently, an interesting atomic displacement pattern is expected for the tetrahedral network of Re ions as schematically depicted in Fig. 4: A Re tetrahedron can expand or shrink alternatingly in the three-dimensional network with keeping the cubic symmetry in total. As a result, a kind of breathing mode is frozen. The modification of the band structure associated with this structural change would be studied in the future work.

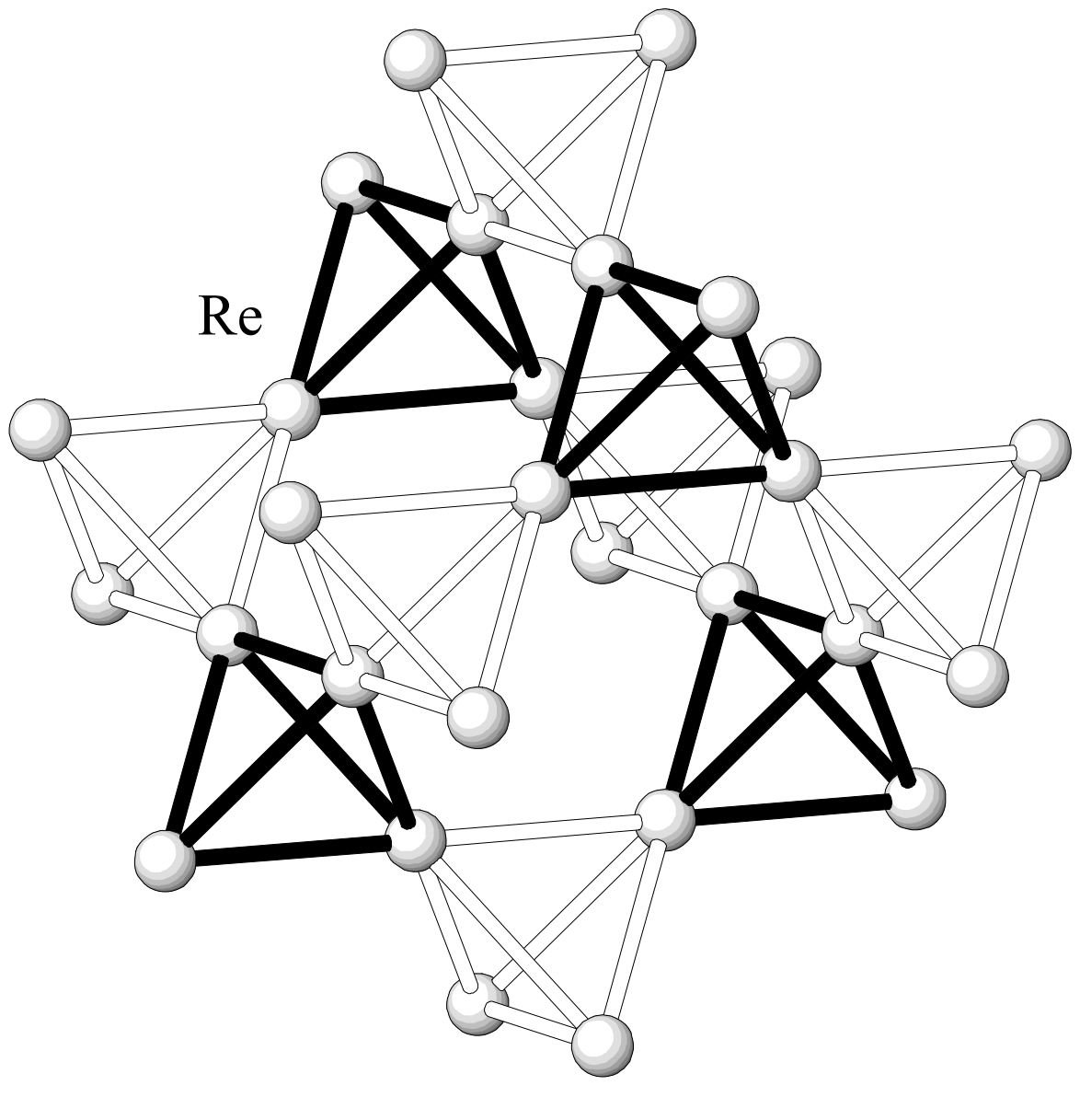

Fig. 4. Tetrahedral network of Re ions in the pyrochlore structure. The Re tetrahedra are identical to each other in the high-temperature structure, while they can expand or shrink alternatingly in the low-temperature structure. Two types of Re tetrahedra are displayed with black and white bonds.

## 4. Concluding Remarks

The pyrochlore oxide superconductor $Cd_2Re_2O_7$ exhibits a novel structural phase transition at $T_s$ = 200 K from the ideal cubic pyrochlore structure (space group $Fd\bar{3}m$) to another cubic structure ($F\bar{4}3m$) on cooling. A large change in electronic structure must accompany the transition, as detected in the resistivity and magnetic susceptibility. It is important to investigate the relation in more detail between the crystal and the electronic structures in the pyrochlore oxides.

A complete structural determination of the low-temperature phase is now in progress using a single crystal XRD. However, we have difficulties in the experiments coming from rather poor crystalline quality due to misoriented domains. Optimization for preparation conditions is necessary to obtain high-quality single crystals.

## Acknowledgments

We thank A. Nakatsuka for his help in making spherical single crystals. This research was supported by Grant-in-Aids for Scientific Research on Priority Areas (A) and for Creative Scientific Research given by The Ministry of Education, Culture, Sports, Science and Technology, Japan.